\documentclass{IEEEtran4PSCC}
\usepackage{graphicx}
\usepackage{threeparttable}
\usepackage{subfigure}
\usepackage{caption}

\usepackage{siunitx}
\DeclareSIUnit[]{\pu}{p.u.}
\DeclareSIUnit[]{\VA}{VA}

\usepackage{colortbl}

\usepackage{cite}

\usepackage[cmex10]{amsmath}

\makeatletter
\let\old@ps@headings\ps@headings
\let\old@ps@IEEEtitlepagestyle\ps@IEEEtitlepagestyle
\def\psccfooter#1{%
    \def\ps@headings{%
        \old@ps@headings%
        \def\@oddfoot{\strut\hfill#1\hfill\strut}%
        \def\@evenfoot{\strut\hfill#1\hfill\strut}%
    }%
    \def\ps@IEEEtitlepagestyle{%
        \old@ps@IEEEtitlepagestyle%
        \def\@oddfoot{\strut\hfill#1\hfill\strut}%
        \def\@evenfoot{\strut\hfill#1\hfill\strut}%
    }%
    \ps@headings%
}
\begin{document}

\title{Dynamic Droop Approach for Storage-based Frequency Control}

%% To specify the authors when (number of affiliations <= 2)
\author{
\IEEEauthorblockN{Yan Jiang\IEEEauthorrefmark{1}, Eliza Cohn\IEEEauthorrefmark{1}, Petr Vorobev\IEEEauthorrefmark{2}, and Enrique Mallada\IEEEauthorrefmark{1}}
\IEEEauthorblockA{\IEEEauthorrefmark{1}Department of Electrical and Computer Engineering, Johns Hopkins University, Baltimore, U.S.A.\\
Email: \{yjiang, ecohn4, mallada\}@jhu.edu}
\IEEEauthorblockA{\IEEEauthorrefmark{2}Center for Energy Science and Technology, Skolkovo Institute of Science and Technology, Moscow, Russia\\
Email: P.Vorobev@skoltech.ru}
}

\maketitle

\begin{abstract}
Transient frequency dips that follow sudden power imbalances --frequency Nadir-- represent a big challenge for frequency stability of low-inertia power systems. Since low inertia is identified as one of the causes for deep frequency Nadir, virtual inertia, which is provided by energy storage units, is said to be one of the solutions to the problem. In the present paper, we propose a new method for frequency control with energy storage systems (ESS),  called dynamic droop control (iDroop), that can completely eliminate frequency Nadir during transients. Nadir elimination allows us to perform frequency stability assessment without the need for direct numerical simulations of system dynamics. We make a direct comparison of our developed strategy with the usual control approaches --virtual inertia (VI) and droop control (DC)-- and show that iDroop is more effective than both in eliminating the Nadir. 
More precisely, iDroop achieves the Nadir elimination under significantly lower gains than virtual inertia and requires almost $40\%$ less storage power capacity to implement the control. Moreover, we show that rather unrealistic control gains are required for virtual inertia in order to achieve Nadir elimination. 
\end{abstract}


% Use this to place sponsorships
%\thanksto{Applicable sponsors, if any, should be placed using the \emph{thanksto} command.}
\thanksto{This work was supported by US DoE EERE award DE-EE0008006, NSF through grants CNS 1544771, EPCN 1711188, AMPS 1736448, and CAREER 1752362, and Johns Hopkins University Discovery Award.}

%----------------------------------------------------%
%               I N T R O D U C T I O N
%----------------------------------------------------%      

\section{Introduction}
The reduction of the systems inertia, which can be caused by replacing conventional synchronous generation with renewable energy sources, is one of the biggest challenges for frequency control in power systems \cite{milano2018}. Lower inertia causes larger transient frequency deviations following sudden power imbalances. Even if the system has adequate primary reserves to keep the steady-state frequency within acceptable limits, low inertia can lead to large transient frequency drops --frequency Nadir--  with unacceptable values. This inability to keep frequency within certain limits is sometimes regarded as the main reason for limiting further increase of renewable penetration. \cite{eirgrid2010all, o2014studying}. Fortunately, the recent advancements in power electronics and electric storage technologies provide the potential to mitigate this degradation through the use of inverter-interfaced storage units that can provide additional frequency response. With proper controllers, fast inverter dynamics can ensure the provision of rapid response from storage devices.

A straightforward --and commonly used-- control approach for energy storage systems (ESS) is to set energy storage units to provide simple proportional power-frequency response, similar to conventional synchronous generators. However, unlike synchronous generators that produce a delayed response to the control signal, the response of storage units is almost instantaneous. This can help arrest the frequency drop during the first few seconds after a disturbance, while generator turbines are gradually increasing their power output. Moreover, because of the absence of delays, smaller droop coefficients (sharper power-frequency response) are accessible for energy storage units, which makes them even more efficient during sudden frequency disturbances. An impressive \SI{200}{\mega\watt} storage service has been procured by the National Grid in the United Kingdom specifically for frequency response with droop coefficient as low as $1\%$ \cite{greenwood2017frequency}. A drawback of such a droop control method is that the energy storage units will continue to provide their response for as long as the system frequency is away from its nominal value, which can lead to rather high requirements of storage capacity.  

Another common control approach is to realize the, so-called virtual inertia service, in which energy storage units imitate the natural inertial response of synchronous machines, thus compensating for the loss of physical inertia of a system. Such a control strategy is especially efficient in reducing the transient frequency drop following a sudden power imbalance. An additional advantage of the virtual inertia is that it only provides response for the duration of frequency transient --usually several seconds, which can significantly reduce the capacity requirements as compared to the droop control strategy. Both virtual inertia (derivative control) and droop (proportional control) can be combined into a single control strategy, and sometimes this combined strategy is referred to as virtual inertia. 

While such a combined service can be very effective in improving the frequency transient performance, the energy storage units have the potential of implementing more complex control strategies. This provides the means for further improvement of the overall frequency response. At the same time, this also reduces the requirements on both power rating and energy capacity of the storage units. In this paper, we show that a novel dynamic droop control strategy -- called iDroop~\cite{m2016cdc,y2017cdc,jiang2019dynamic} -- outperforms both droop control and virtual inertia, while still being a rather simple first-order control. We demonstrate that by properly choosing the iDroop parameters, it is possible to completely eliminate the frequency Nadir. In fact, the frequency dynamics become first-order, so that the response to a sudden imbalance is monotonic. Thus, the system  frequency steadily moves towards its final equilibrium value, which is determined by the available primary reserves. In addition to such an outstanding transient performance, the iDroop control also reduces the required power rating and energy capacity of the storage units, making it a much more efficient control strategy.  

The structure of the manuscript is as follows. In Section \ref{Sec:modelling}, we present the modelling approach and discuss the main challenges to low-inertia frequency control. Section \ref{Sec:Virt} is dedicated to the analysis of simple control approaches, namely droop control and virtual inertia. Next, in Section \ref{Sec:iDroop}, the  iDroop control strategy is presented and its benefits are discussed in detail. The results are summarized and future research directions are discussed in Section \ref{Sec:conc}.  
 %\emph{how-to} 

%----------------------------------------------------%
%               S E C T I O N  II
%----------------------------------------------------%      

\section{Modelling Approach}\label{Sec:modelling}
% \begin{enumerate}
%     \item Introduce a one-bus model without storage, show nadir, preferrably with real-life values
    
%   \item Discuss the metrics for storage control
% \end{enumerate}
In this section we establish the basic model of a power system that we use to study the dynamic performance of different frequency control strategies. Since the issue of big frequency deviations is most apparent for compact power systems (including microgrids)\cite{o2014studying}, it is sufficient for the purposes of the present manuscript to consider a single-area approximation \cite{lalor2005impact} that corresponds to the dynamics of the system center of inertia (COI). Throughout the manuscript we will use the small letters to denote the deviation of a corresponding variable from its nominal value; for example $\omega= \Omega - \Omega_0$ is the frequency deviation from the nominal frequency ($\Omega_0=60$ Hz in this manuscript). Next, we use hat to denote variables in the Laplace domain, e.g. $\hat{\omega}(s)$ is the Laplace transform of the frequency deviation $\omega$.   

\begin{figure}[!t]
\centering
\includegraphics[width=0.85\columnwidth]{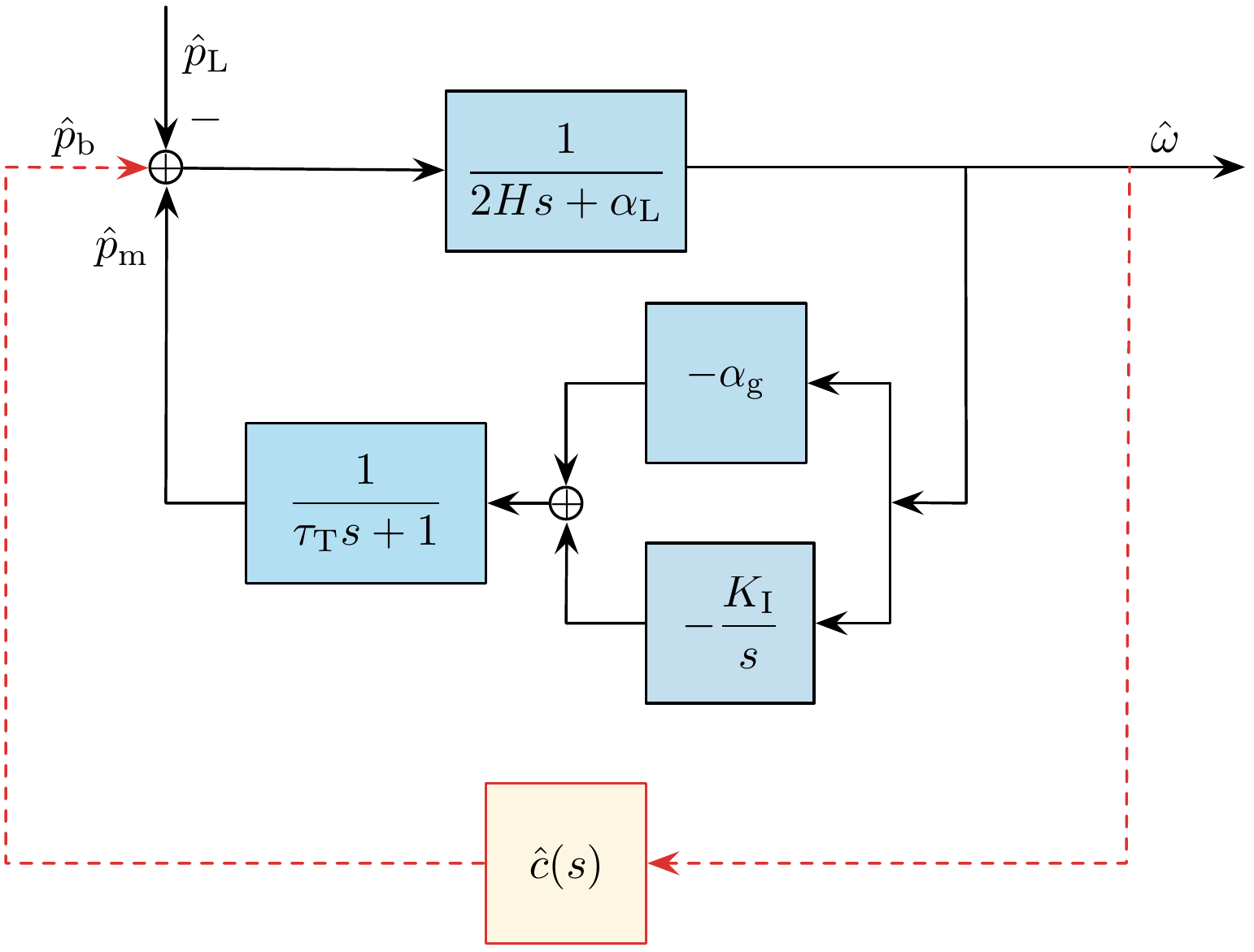}
\caption{Block diagram of an aggregated power system with frequency control from generators and storage units.}
\label{fig:1bus-model}
\end{figure}

The block-diagram (in the Laplace domain) of an aggregated power system together with frequency control blocks is shown in Fig.~\ref{fig:1bus-model}. The system consists of an equivalent synchronous machine with a turbine that has both primary (proportional) and secondary (integral) frequency control loops. An additional frequency control with transfer function $\hat{c}(s)$ is realized by a storage unit. The particular form of $\hat{c}(s)$ depends on the control strategy. The input to the system of Fig.~\ref{fig:1bus-model} is a power imbalance $\hat{p}_\mathrm{L}$ which can be the result of either load or generation variation. The output of the system is $\hat{\omega}$ - the frequency deviation from the nominal value. 

The dynamic equations for the system in Fig.~\ref{fig:1bus-model} in time domain have the following form: 
% \begin{subequations}\label{eq:dyn-onebus}
% \begin{align}
% 2H \dot{\omega} =&\ p_\mathrm{m} -p_\mathrm{L} - \alpha_\mathrm{L} \omega+p_\mathrm{b}\;,\\
%     \tau_\mathrm{T} \dot{p}_\mathrm{m} =& -p_\mathrm{m} - \alpha_\mathrm{g} \omega\;,\label{eq:turbine-dyn} 
% \end{align}
%\end{subequations}
\begin{subequations}\label{eq:dyn-onebus}
\begin{align}
    \dot{\theta} =&\ \omega\;,\\
    2H \dot{\omega} =&\ p_\mathrm{m} -p_\mathrm{L} - \alpha_\mathrm{L} \omega+p_\mathrm{b}\;,\\
    \tau_\mathrm{T} \dot{p}_\mathrm{m} =& -p_\mathrm{m} - \alpha_\mathrm{g} \omega - K_\mathrm{I}\theta\;,\label{eq:tur-dyan}\\
    \dot{E}_\mathrm{b} =&\ p_\mathrm{b}\;,
\end{align}
\end{subequations}
where the parameters are defined as: $H$ - the inertia time constant of the system, $\tau_\mathrm{T}$ - the turbine time constant, $\alpha_\mathrm{L}$ - the load-frequency sensitivity coefficient, $\alpha_\mathrm{g}$  - the aggregate inverse droop of generators, and $K_\mathrm{I}$ - the secondary frequency control gain. The state variables are: $\omega$ - frequency deviation from the nominal value, $\theta$ - the integral of the frequency deviation, $p_\mathrm{m}$ - the per unit turbine power deviation from the nominal value, and $E_\mathrm{b}$ - the energy supplied from the storage unit. $p_\mathrm{b}$ is the storage unit power output. 

We use the data from the Great Britain power system as a reference in this manuscript, with values taken mostly from \cite{GBoper2016} and \cite{GBpower}. According to \cite{GBoper2016}, the value of inertia in $2025$ under the high renewable penetration scenario will be as low as \SI{70}{\giga\VA\second}, which corresponds to an inertia constant of $H=\SI{2.19}{\second}$ on a \SI{32}{\giga\watt} basis. The maximum value of a sudden power imbalance that the system should survive is $\Delta P=\SI{1.8}{\giga\watt}$ according to \cite{GBpower}. Also, we assume that the aggregate inverse droop of the generators is $\alpha_\mathrm{g}=\SI{15}{\pu}$ and the load sensitivity coefficient is $\alpha_\mathrm{L}=\SI{1}{\pu}$. The data is summarized in Table~\ref{table:parameter}.     
%\cite{Vorobev2019tps}

\begin{figure}[!t]
\centering
\includegraphics[width=.8\columnwidth]{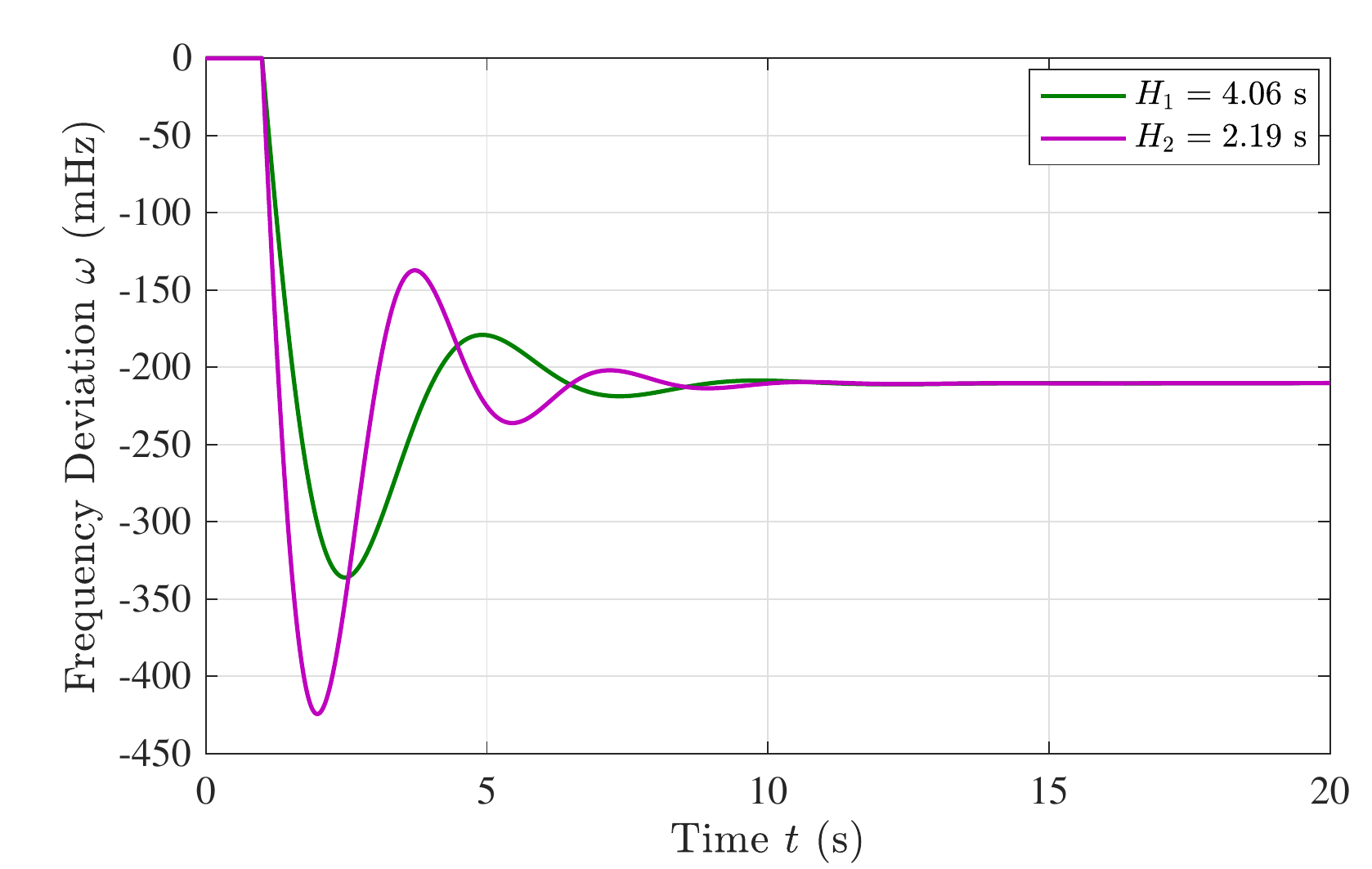}
\caption{Frequency response to a sudden loss of a generator unit for two different values of the total system inertia with primary control only from generators.}
\label{fig:fre-sw}
\end{figure}

\begin{table}[!ht]
\renewcommand{\arraystretch}{1.3}
\centering
\caption{Values of Power Systems Parameters}
\begin{threeparttable}[t]
\begin{tabular}{c|c|c}
\arrayrulecolor[rgb]{.7,.8,.9}
\hline\hline
Parameters & Symbol & Value\\
\hline
System power base & $P_\mathrm{B}$ & \SI{32}{\giga\watt} \\
\hline
Maximum power imbalance & $\Delta P$ & \SI{1.8}{\giga\watt} \\
\hline
Inertia time constant 
% \footnotemark[5] 
& $H$ & \SI{2.19}{\second} \\
\hline
Turbine time constant & $\tau_\mathrm{T}$ & \SI{1}{\second}\\
\hline
Load sensitivity coefficient & $\alpha_\mathrm{L}$ & \SI{1}{\pu} \\ 
\hline
Aggregate inverse droop of generators & $\alpha_\mathrm{g}$ & \SI{15}{\pu}\\
\hline
Secondary frequency control gain & $K_\mathrm{I}$ & \SI{0.05}{\per\second}\\
\hline\hline
\end{tabular}
\begin{tablenotes}
     \item[1] All per unit values are on the system power base.
\end{tablenotes}
\end{threeparttable}
\label{table:parameter}
\end{table}

To offer a better insight of the importance of introducing inverter-interfaced storage for improving frequency transient performance in low-inertia power systems, we provide a plot of frequency deviations following a power disturbance $p_\mathrm{L}$. Fig.~\ref{fig:fre-sw} illustrates the responses of the system frequency to a sudden power imbalance (typically, a loss of a generating unit) for two different values of system inertia with droop control executed only by generators. The values of inertia are chosen according to \cite{GBoper2016} - $H_1=\SI{4.06}{\second}$ corresponds to the present lowest value of the system inertia in Great Britain, while $H_2=\SI{2.19}{\second}$ corresponds to the expected inertia in $2025$ under the high renewable penetration scenario. Clearly, for the lower value of inertia, the transient frequency dip is unacceptably low. Thus, certain measures should be taken in order to reduce it.

A widely embraced approach to mitigate this problem is to employ inverter-interfaced storage for frequency control\cite{knap2015sizing}. Various control strategies can be used to provide the storage output $p_\mathrm{b}$ as a function of the frequency deviation $\omega$. In the Laplace domain, the power response to frequency variation can be described as  
\begin{equation}\label{eq:inverter-tf}
\hat{p}_\mathrm{b}(s)=\hat{c}(s)\hat \omega(s)\,,
\end{equation}
with $\hat{c}(s)$ specified by the transfer function of the corresponding control strategy.

% With only the primary frequency control, a load perturbation usually results in a nonzero steady-state frequency deviation. Hence, in order to restore the frequency to its nominal value, we need to introduce the secondary frequency control to the one-area model by adding an integral term to the right hand side of \eqref{eq:turbine-dyn}. Then the (partial) system dynamic equations become
% \begin{subequations}\label{eq:dyn-onebus-2nd}
% \begin{align}
%     \dot{\theta} =&\ \omega\;,\\
%     2H \dot{\omega} =&\ p_\mathrm{m} -p_\mathrm{L} - \alpha_\mathrm{L} \omega+p_\mathrm{b}\;,\\
%     \tau_\mathrm{T} \dot{p}_\mathrm{m} =& -p_\mathrm{m} - \alpha_\mathrm{g} \omega - K_\mathrm{I}\theta\;,\\
%     \dot{E}_\mathrm{b} =&\ p_\mathrm{b}\;,
% \end{align}
% \end{subequations}
% where $\theta$ is the integral of frequency deviation and $K_\mathrm{I}$ the secondary frequency control gain with a typical value \SI{0.05}{\per\second}. Here, the energy supply $E_\mathrm{b}$ from the inverter-interfaced storage is also of our interest.

When designing frequency control strategies by energy storage, constraints come not only from the system performance, but also from economic factors. Therefore, power and energy capacity requirements, the two main parameters that largely affect the economy of storage units, should also be considered as performance metrics when developing different control strategies. We will define these two metrics as follows:
\begin{itemize}
\item Power requirement is the maximum amount of power output from storage during the whole transient duration, which can be quantified as:
\[
p_\mathrm{b,max} = \max_{t\geq0} p_\mathrm{b}(t)\,;
\]
\item Energy capacity is the maximum amount of energy supply from storage during the whole transient duration, which can be quantified as:
\[
E_\mathrm{b,max}= \max_{t\geq0} E_\mathrm{b}(t)\,.
\]
\end{itemize}
In order to make the analysis tractable, we assume $K_\mathrm{I}=0$ in \eqref{eq:dyn-onebus} whenever evaluating $p_\mathrm{b,max}$ analytically since secondary frequency control is typically very slow (around $10$ minutes) and does not have a significant impact on the transient frequency dynamics in the first several seconds after the disturbance, which is where frequency Nadir is observed.

%----------------------------------------------------%
%               S E C T I O N  III
%----------------------------------------------------%  

\section{Inverter-interfaced Storage for Frequency Control}\label{Sec:Virt}
% \begin{enumerate}
%  \item Show typical storage options: DC, VI
    
%     \item Show the nadir-less control with VI, calculate the required amount of power
    
%     \item Show the disadvantages of DC (a lot of storage capacity) and VI, explain the logic behind the dynamic droop
% \end{enumerate}
In this section we briefly analyze the traditional control strategies for storage -- droop control and virtual inertia. We assess the performance of these two strategies from the point of view of the improvement of transient frequency dynamics as well as the required amounts of power and energy from storage to improve the frequency Nadir.

%\subsection{Typical Control Strategies on Inverter-interfaced Storage}

The most common control strategy for inverter-interfaced storage combines both droop control and virtual inertia and can be represented by the following effective storage transfer function $\hat{c}_\mathrm{vi}(s)$:
\begin{align}\label{vimain}
    \hat{c}_\mathrm{vi}(s):=-(m_\mathrm{v}s+\alpha_\mathrm{b})\;,
\end{align}
where $\alpha_\mathrm{b}$ is the inverse storage droop and $m_\mathrm{v}$ the virtual inertia constant. Notably, if $m_\mathrm{v}=0$, then virtual inertia reduces to droop control, i.e.,
\begin{align}
    \hat{c}_\mathrm{dc}(s):=-\alpha_\mathrm{b}\;,
\end{align}
which only provides additional droop capability. It is common in the literature to refer to the control strategy given by \eqref{vimain} simply as "virtual inertia", hence our subscript "vi" used for $\hat{c}$. We will also use this convention whenever needed.  

% \begin{figure}[!t]
% \centering
% \includegraphics[width=0.6\columnwidth]{Figures/inverter_VI}
% \caption{Block diagram of VI.}
% \label{fig:VI}
% \end{figure}

\subsection{Nadir Elimination via Virtual Inertia}

\begin{figure}[t!]
\centering
\includegraphics[width=0.8\columnwidth]{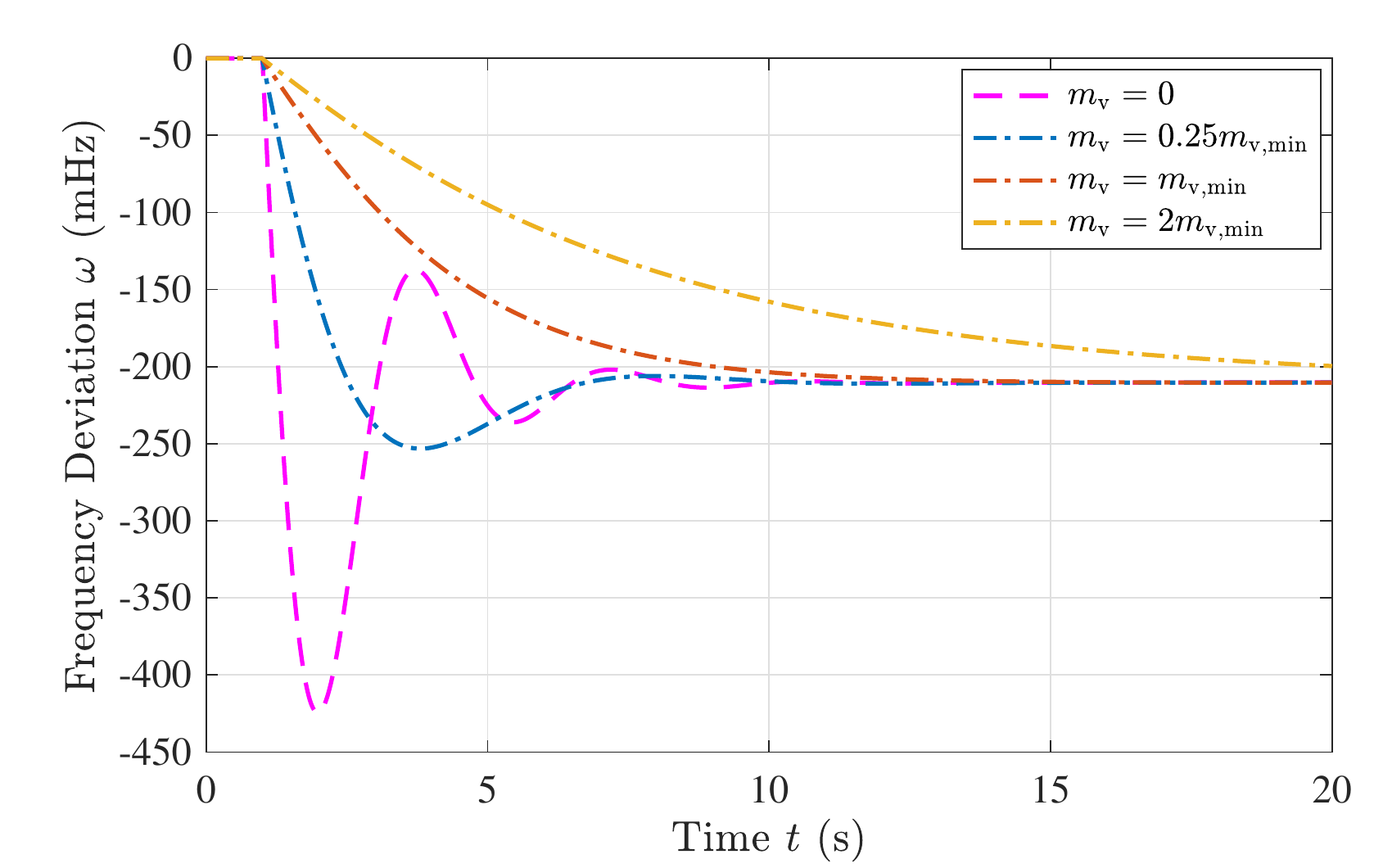}
\caption{Frequency deviations under virtual inertia control from storage with $\alpha_\mathrm{b}=0$ and different values of  $m_{\mathrm{v}}$.}
\label{fig:fre-vi}
\end{figure}

Fig.~\ref{fig:fre-vi} shows the frequency dynamics following a step power imbalance of $\Delta P = \SI{1.8}{\giga\watt}$, for the system described by Fig.~\ref{fig:1bus-model}, under virtual inertia control from storage with $\alpha_\mathrm{b}=0$ and different values of  $m_{\mathrm{v}}$. Note that although the steady-state frequency deviation is always determined by the aggregate inverse droop of the system as
\begin{align}\label{eq:fre-ss}
\omega_\mathrm{vi}(\infty)=-\frac{\Delta P}{\alpha_\mathrm{L}+\alpha_\mathrm{g}+\alpha_\mathrm{b}},
\end{align}
the frequency Nadir significantly depends on the choice of the storage inertia constant $m_{\mathrm{v}}$. Obviously, one can entirely remove the Nadir without affecting the steady-state frequency deviation by tuning $m_{\mathrm{v}}$ appropriately. 

In fact, to achieve Nadir elimination, the storage control parameters $(\alpha_\mathrm{b}, m_{\mathrm{v}})$ should satisfy the following relation \cite{jiang2019dynamic}:
\begin{align}\label{eq:nadir-VI}
     \alpha_\mathrm{b} \leq  \left(2H+ m_{\mathrm{v}}\right)\left(\tau_\mathrm{T}^{-1} - 2\sqrt{\frac{\tau_\mathrm{T}^{-1}\alpha_\mathrm{g}}{2H+ m_{\mathrm{v}}}}\right)-\alpha_\mathrm{L}\;.
\end{align}
This condition can be better understood considering the case with  $m_\mathrm{v}=0$, i.e., pure droop control from the storage. In this case, a small $H$ inherited by a low-inertia power system can result in a negative right hand side in condition \eqref{eq:nadir-VI}, which means that there exists no feasible $\alpha_\mathrm{b}$ for Nadir elimination. Therefore, droop control alone is unable to eliminate Nadir and some amount of storage inertia $m_\mathrm{v}$ is needed. 
It follows from \eqref{eq:nadir-VI} that, given $\alpha_\mathrm{b}$, the amount of $m_{\mathrm{v}}$ needed to eliminate Nadir must be sufficiently large, i.e.,
\begin{align}\label{eq:mvmin}
   m_{\mathrm{v}} \geq m_\mathrm{v,min} =\tau_\mathrm{T}\beta^2-2H\,, %\tau_\mathrm{T}\left[\alpha_\mathrm{L}+\alpha_\mathrm{b}+2\alpha_\mathrm{g}+2\sqrt{\alpha_\mathrm{g}\left(\alpha_\mathrm{L}+\alpha_\mathrm{b}+\alpha_\mathrm{g}\right)}\right]-2H.
\end{align}
with
\[
\beta=\sqrt{\alpha_\mathrm{g}}+\sqrt{\alpha_\mathrm{L}+\alpha_\mathrm{g}+\alpha_\mathrm{b}}\;.
\]
If it is recognized that both $\alpha_\mathrm{b}$ and $\alpha_\mathrm{L}$ are much less than $\alpha_\mathrm{g}$, a linear approximation of the expression for $m_\mathrm{v,min}$ in \eqref{eq:mvmin} can be used without much loss of accuracy:
\begin{align}\label{eq:mvmin-linear}
   m_\mathrm{v,min} \approx a \alpha_\mathrm{b} + b\,,
\end{align}
where $a=2\tau_\mathrm{T}$ and $b=4 \tau_\mathrm{T}\alpha_\mathrm{g}-2H$.
% \begin{align*}
    % a=&\ 2\tau_\mathrm{T}\;,\\
    % b=&\ 4 \tau_\mathrm{T}\alpha_\mathrm{g}-2H\;.
% \end{align*}
% \begin{align*}
%     a:=&\ 2\tau_\mathrm{T}\;,\\
%     b:=&\ \tau_\mathrm{T}\left(\alpha_\mathrm{L}+4\alpha_\mathrm{g}\right)-2H\;.
% \end{align*}
% \begin{align*}
%     a:=&\ \tau_\mathrm{T}\left(1+\sqrt{\frac{\alpha_\mathrm{g}}{\alpha_\mathrm{L}+\alpha_\mathrm{g}}}\right)\;,\\
%     b:=&\ \tau_\mathrm{T}\left[\alpha_\mathrm{L}+2\alpha_\mathrm{g}+2\sqrt{\alpha_\mathrm{g}\left(\alpha_\mathrm{L}+\alpha_\mathrm{g}\right)}\right]-2H\;.
% \end{align*}
The importance of this linear approximation is that it represents a closed-form expression that can be used to directly calculate the needed storage inertia constant $m_\mathrm{v,min}$ required to eliminate Nadir, once inverse storage droop $\alpha_\mathrm{b}$ is determined. The latter can be found from \eqref{eq:fre-ss} by demanding some steady-state frequency deviation once the size of the maximum power disturbance is specified. Suppose the expected maximum magnitude of power imbalance is $\Delta P$ and the demanded maximum frequency deviation is $\Delta \omega$. Then one needs to choose
\begin{align}\label{eq:ab-design}
\alpha_\mathrm{b}=\left|\frac{\Delta P}{\Delta \omega}\right| - \alpha_\mathrm{g}\,.
\end{align}
After that, the required storage inertia constant can be found from \eqref{eq:mvmin-linear}:
\begin{align}\label{eq:mvmin-exp}
m_\mathrm{v,min} = 2\tau_T \left|\frac{\Delta P}{\Delta \omega}\right|+2\tau_T \alpha_\mathrm{g} - 2H\,.
\end{align}
In the case where the right-hand side of \eqref{eq:ab-design} returns negative, $\alpha_\mathrm{b}$ can be set to zero with the corresponding change in \eqref{eq:mvmin-exp}.

The significance of the above derivations lies in that the frequency security of the system was certified by performing only \emph{algebraic} calculations, without the need to run explicit \emph{dynamic} simulations.

\begin{figure}[t!]
\centering
\includegraphics[width=.8\columnwidth]{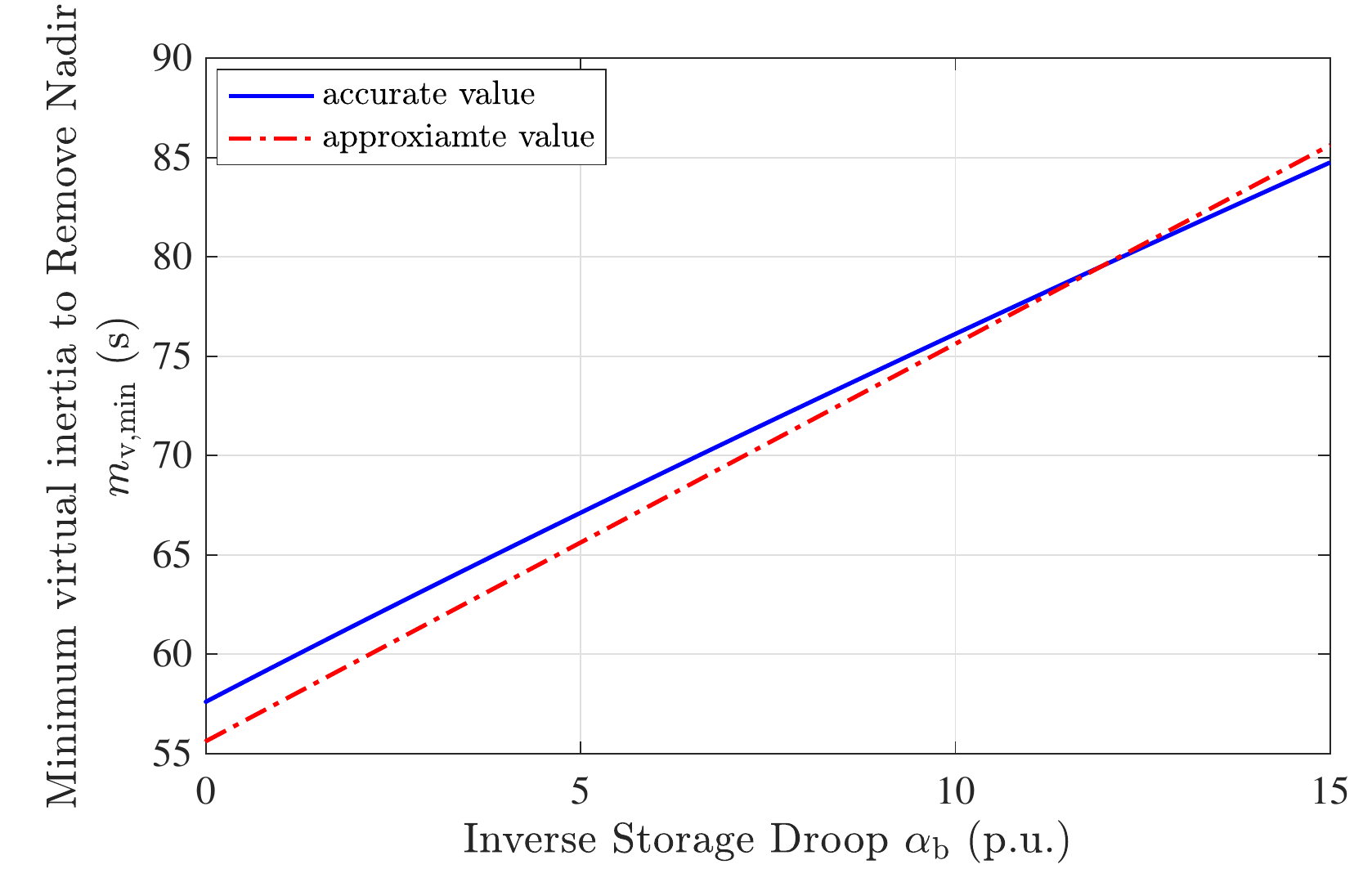}
\caption{Minimum virtual inertia constant required to remove Nadir as $\alpha_\mathrm{b}$ varies within $0$ to \SI{15}{\pu}.}
\label{fig:mvmin}
\end{figure}

As an illustration, Fig.~\ref{fig:mvmin} shows the minimum virtual inertia constant  requirement $m_\mathrm{v,min}$ for eliminating the Nadir as a function of the inverse storage droop $\alpha_\mathrm{b}$. Both the exact solution from \eqref{eq:mvmin} and the linear approximation from \eqref{eq:mvmin-linear} are shown, thus demonstrating the minimal difference between the two. One thing to note is that the $m_\mathrm{v,min}$ required has rather high values, the equivalent of more than $30$ times of the actual system inertia (under the high renewable penetration scenario). This leads to a very long settling time, as seen from the Fig.~\ref{fig:fre-vi}. More importantly, high gain from a derivative term makes the whole control strategy very sensitive to measurement noise and delays. The detailed analysis of this issue is beyond the scope of the present manuscript and is the subject of subsequent publications.

\subsection{Power and Energy Capacity Required by Virtual Inertia}

\begin{figure}[t!]
\centering
\subfigure[Maximum frequency deviation]
{\includegraphics[width=0.8\columnwidth]{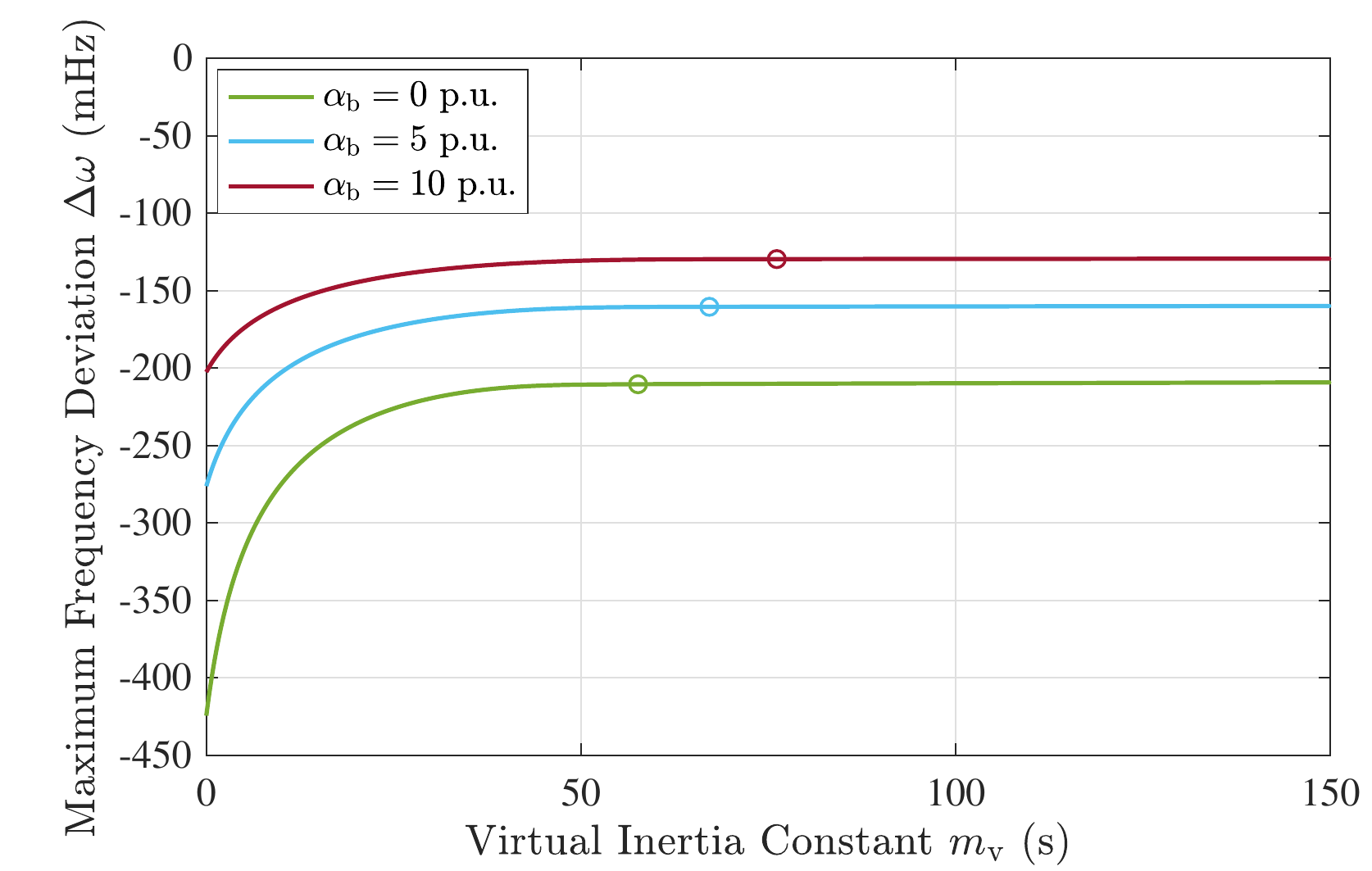}}
\subfigure[Power capacity]
{\includegraphics[width=0.8\columnwidth]{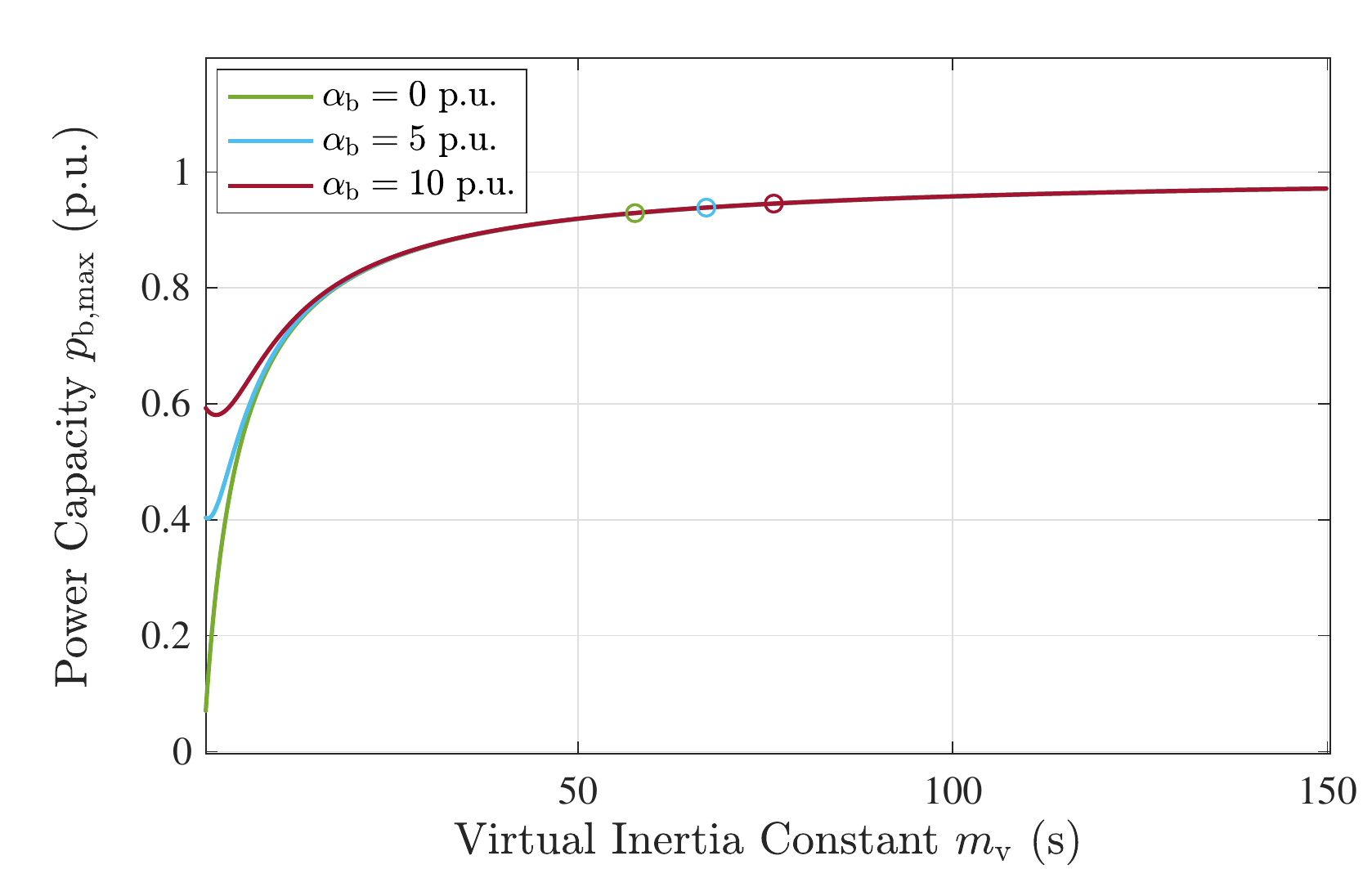}}
\caption{Effect of virtual inertia constant on maximum frequency deviation and power capacity requirement, where circles denote the points corresponding to $m_\mathrm{v}=m_\mathrm{v,min}$.}
\label{fig:saturation-vi}
\end{figure}

To provide a better understanding of the role of the virtual inertia storage constant in Nadir elimination, Fig.~\ref{fig:saturation-vi} shows the effect of $m_\mathrm{v}$ on the maximum frequency deviation $\Delta\omega$ and the required storage power capacity $p_\mathrm{b,max}$ for different chosen values of $\alpha_\mathrm{b}$. \footnote{ For the purpose of fair comparison, in this paper, we refer the required storage power and energy capacity to the value of the power disturbance, not the system base power.} Clearly, for $m_\mathrm{v}$ less than $m_\mathrm{v,min}$, frequency deviation and power capacity are extremely sensitive to variations in $m_\mathrm{v}$, yet, for $m_\mathrm{v}$ greater than $m_\mathrm{v,min}$, they are practically insusceptible to changes in $m_\mathrm{v}$. This implies that $m_\mathrm{v}=m_\mathrm{v,min}$ plays the role of a saturation point after which an increase in power capacity does not provide any benefit to a decrease in frequency deviation. This justifies the optimality of Nadir elimination by setting $m_\mathrm{v}=m_\mathrm{v,min}$. Such a choice makes the system critically damped. 

The storage energy capacity required to execute the virtual inertia and droop control is mostly dominated by the values of $\alpha_\mathrm{b}$ used. For values of $\alpha_\mathrm{b}$ that are not very small, it can be calculated as $E_{\mathrm{b,max}} = \alpha_\mathrm{b}/K_\mathrm{I}$. This suggests that higher secondary control gains tend to reduce the required storage energy capacity.

\section{Dynamic Droop Control}\label{Sec:iDroop}
% \begin{enumerate}
%     \item Show the intuition behind dynamic droop
    
%     \item Nadir-less control
    
%     \item Power and capacity requirements
    
%     \item What if we don't know the turbine time constant exactly?
    
%     \item What if turbine is of higher order (with some fast governor)?
% \end{enumerate}
 In this section we turn to an alternative control strategy, the iDroop control~\cite{m2016cdc}, which is able to eliminate the Nadir and overcome the drawbacks of virtual inertia discussed in the previous section. In order to give the intuition behind the iDroop control strategy, we first note that inverter-interfaced storage units are potentially capable of executing a much wider class of control strategies than droop and/or virtual inertia. The idea behind the iDroop approach is to provide an enhanced response to frequency, only during the transient, in order to eliminate the Nadir. Then, once the generator turbine is in action, we withdraw the excessive control effort. This can be achieved by combining the proportional and the lag element in a single transfer function. 
% \enrique{At this point I kind of feel that energy requirements should be minimally discussed... the title says Energy capacity, and we always talk about it, but it is not dicussed here.}
 
 \subsection{Nadir Elimination via Dynamic Droop}
\begin{figure}[!t]
\centering
\includegraphics[width=0.8\columnwidth]{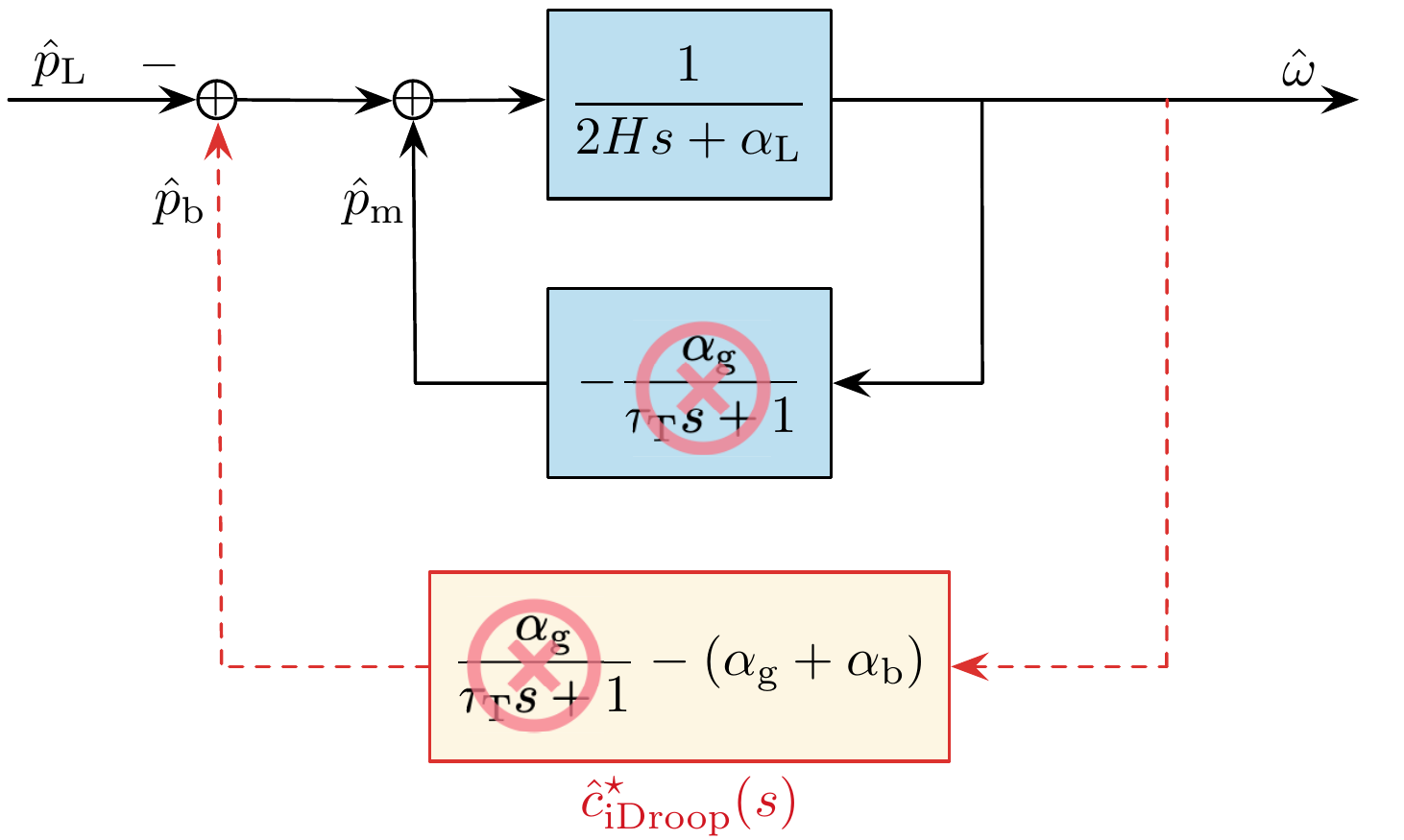}
\caption{Cancellation of turbine dynamics using iDroop.}
\label{fig:idroop-block}
\end{figure}

\begin{figure}[t!]
\centering
\includegraphics[width=0.8\columnwidth]{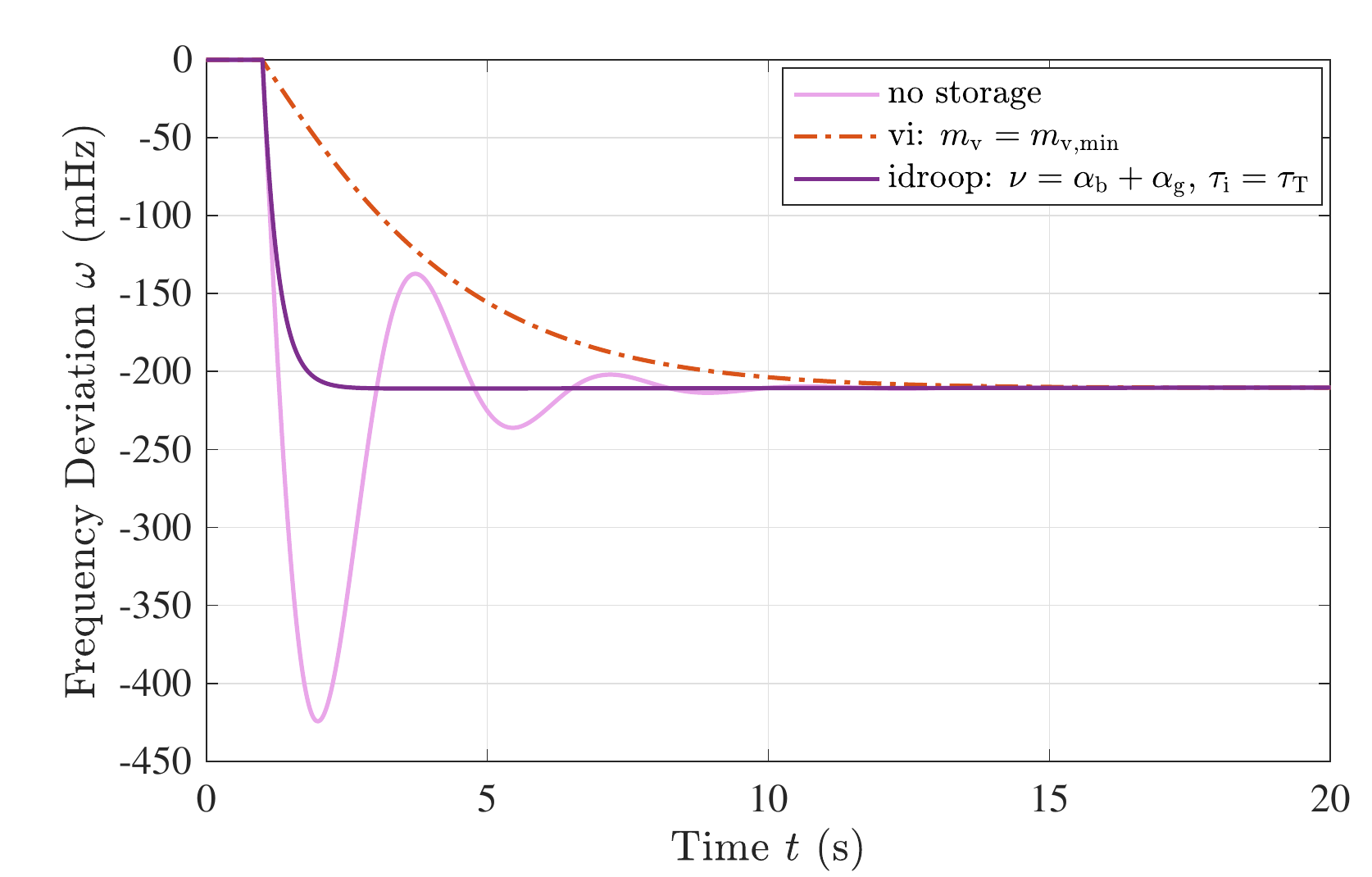}
\caption{Comparison of frequency deviations under virtual inertia and iDroop with Nadir elimination tuning to a step
power imbalance for $\alpha_\mathrm{b} = 0$.}
\label{fig:dyn-fre-idroop-ab0_nodb}
\end{figure}

 The endorsed control strategy can be described by the transfer function:
\begin{align}\label{eq:tf-idroop}
    \hat{c}_\mathrm{idroop}(s)=\frac{\nu- \alpha_\mathrm{b}}{\tau_\mathrm{i} s+1}-\nu\;,
\end{align}
where $\nu>0$ and $\tau_\mathrm{i}>0$ are the tunable parameters. As it was pointed out before, the virtue of iDroop is that it is composed of a lag element in parallel with a proportional element, of which the former can be used to cancel out the turbine dynamics and the latter can be used to improve the steady-state frequency deviation. This is exactly the intuition behind the Nadir elimination tuning proposed in \cite{jiang2019dynamic}. More precisely, by tuning 
\begin{equation}
\nu=\alpha_\mathrm{b}+\alpha_\mathrm{g}\qquad\text{and}\qquad\tau_\mathrm{i}={\tau_\mathrm{T}}\,,
\end{equation}
iDroop can successfully eliminate the Nadir since the transfer function of iDroop turns into the form of
\begin{equation}\label{eq:co-idroop-nonadir}
\hat{c}^\star_\mathrm{idroop}(s) =\frac{ \alpha_\mathrm{g} }{\tau_\mathrm{T} s+1}-\left(\alpha_\mathrm{g} + \alpha_\mathrm{b}\right)\;,
\end{equation}
where the first term above cancels out the turbine dynamics (Fig.~\ref{fig:idroop-block}) and thus makes the system effectively first-order. This enables the frequency to evolve monotonously towards its steady-state value in response to a step power perturbation. Fig.~\ref{fig:dyn-fre-idroop-ab0_nodb} illustrates this by showing the dynamics of the frequency following a step power perturbation for the case of no storage, virtual inertia, and iDroop control from storage. 

%\subsection{Storage Required by Dynamic Droop in Nadir Elimination }

\subsection{Power and Energy Capacity Required by iDroop}

\begin{figure}[t!]
\centering
\subfigure[Power output from turbine]
{\includegraphics[width=0.8\columnwidth]{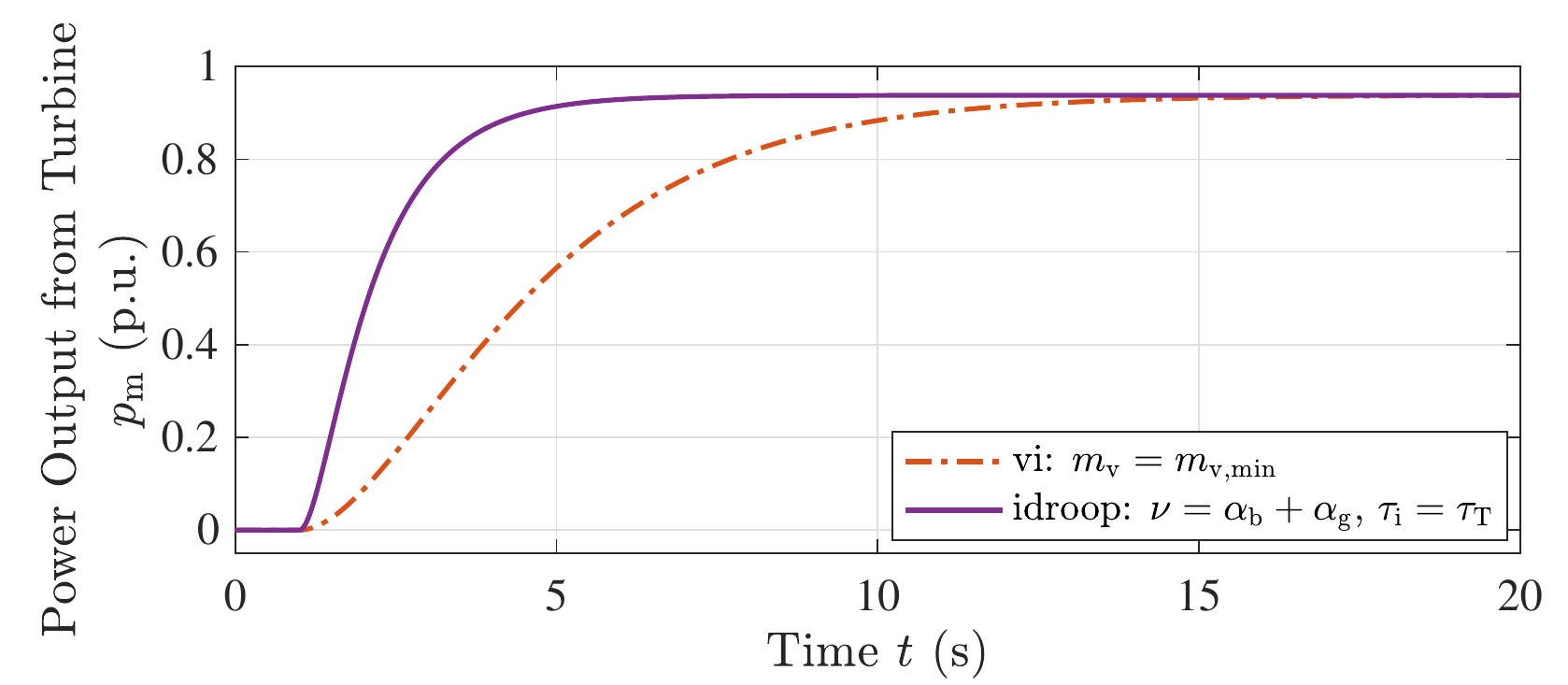}}\label{fig:dyn-Pm-idroop-ab0_nodb}
\subfigure[Power output from storage]
{\includegraphics[width=0.8\columnwidth]{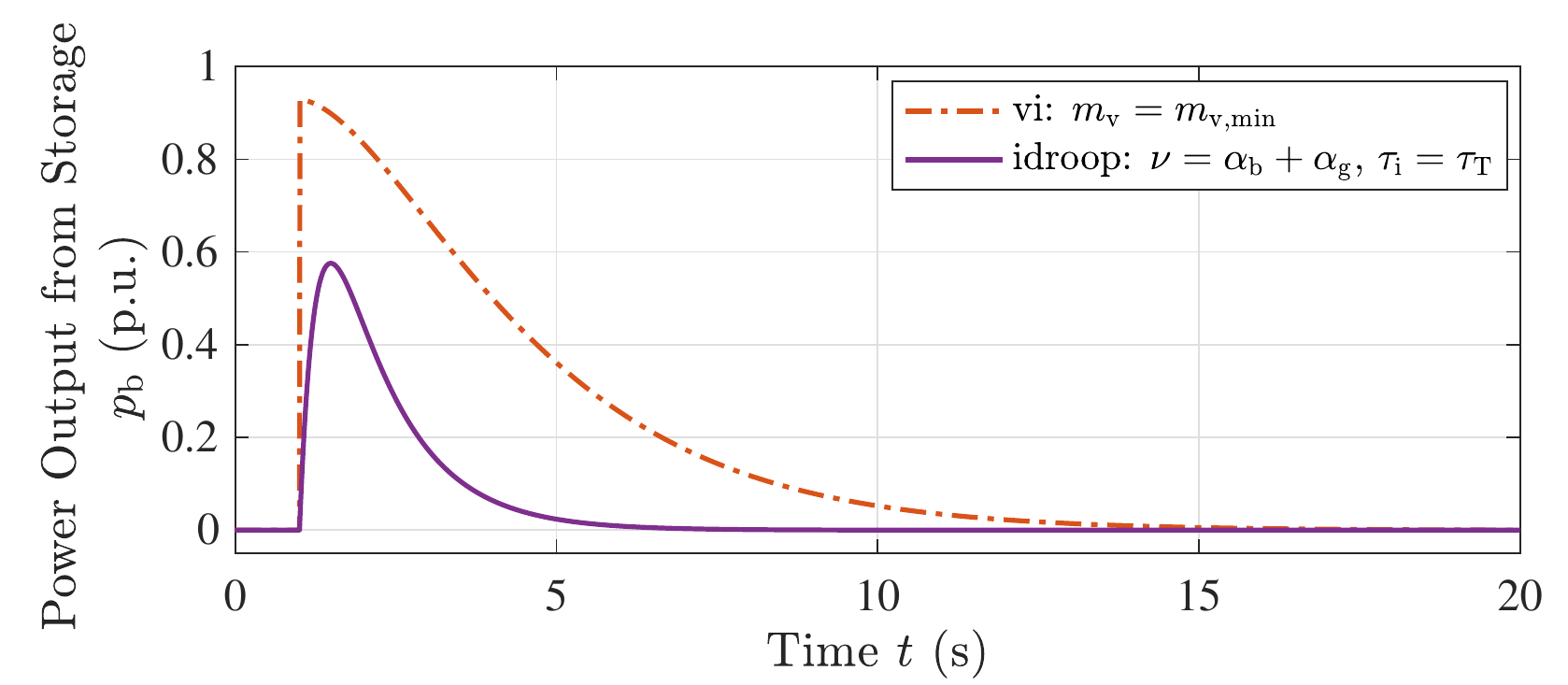}}\label{fig:dyn-Pb-idroop-ab0_nodb}
\subfigure[Energy supply from storage]
{\includegraphics[width=0.8\columnwidth]{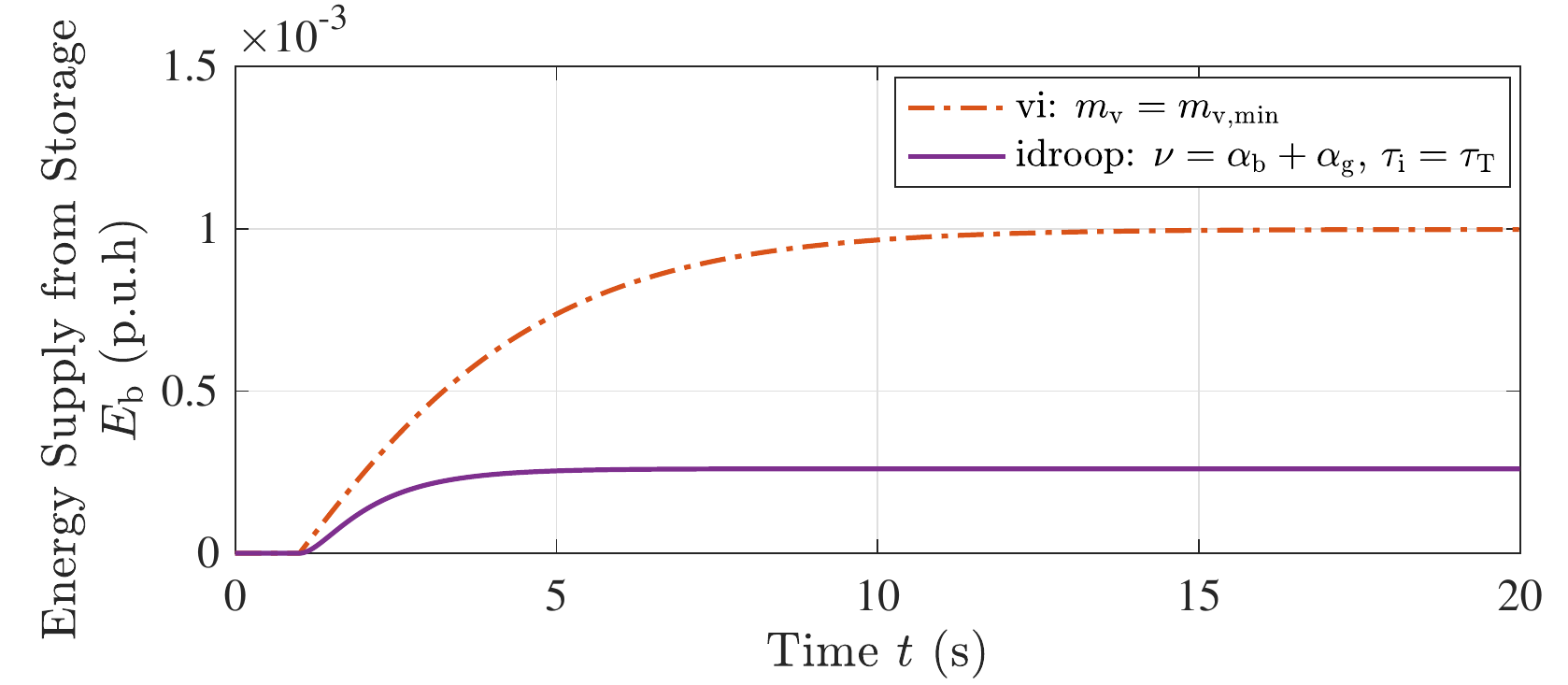}}\label{fig:dyn-Eb-idroop-ab0-nodb}
\caption{Comparison of power and energy transient responses under virtual inertia and iDroop with Nadir elimination tuning for a step
power imbalance for $\alpha_\mathrm{b} = 0$.}
\label{fig:dyn-nodb-ab0}
\end{figure}

We show in Fig.~\ref{fig:dyn-nodb-ab0} the dynamics of the storage power output during the transient for two different control strategies: virtual inertia \eqref{vimain} and iDroop \eqref{eq:co-idroop-nonadir}, both with $\alpha_\mathrm{b}=0$. To avoid confusion, we note that the storage power in Fig.~\ref{fig:dyn-nodb-ab0} is relative to the magnitude of the power disturbance, instead of the system base power. It is clear that the iDroop approach allows us to eliminate the Nadir with much less storage power required than the virtual inertia approach. The additional advantage lies in the fact that the time interval during which the storage is active is significantly shorter for iDroop than that for virtual inertia. The physical intuition behind this is that iDroop provides a high proportional response initially, and then gradually withdraws its participation, letting the  turbine pick up the total imbalance in the long run. Thus, we can say that iDroop takes the full advantage of the system internal control capabilities, providing just enough additional control efforts to eliminate the frequency Nadir. We also note that, as evident from Fig.~\ref{fig:dyn-fre-idroop-ab0_nodb}, the initial rate of change of frequency (RoCoF) is higher in the case of iDroop approach than that of the simple virtual inertia control. This can be addressed separately, if the limits on RoCoF are present, and will require some additional tuning of the iDroop settings. We will address this issue in subsequent publications.  

\begin{figure}[!t]
\centering
\includegraphics[width=.85\columnwidth]{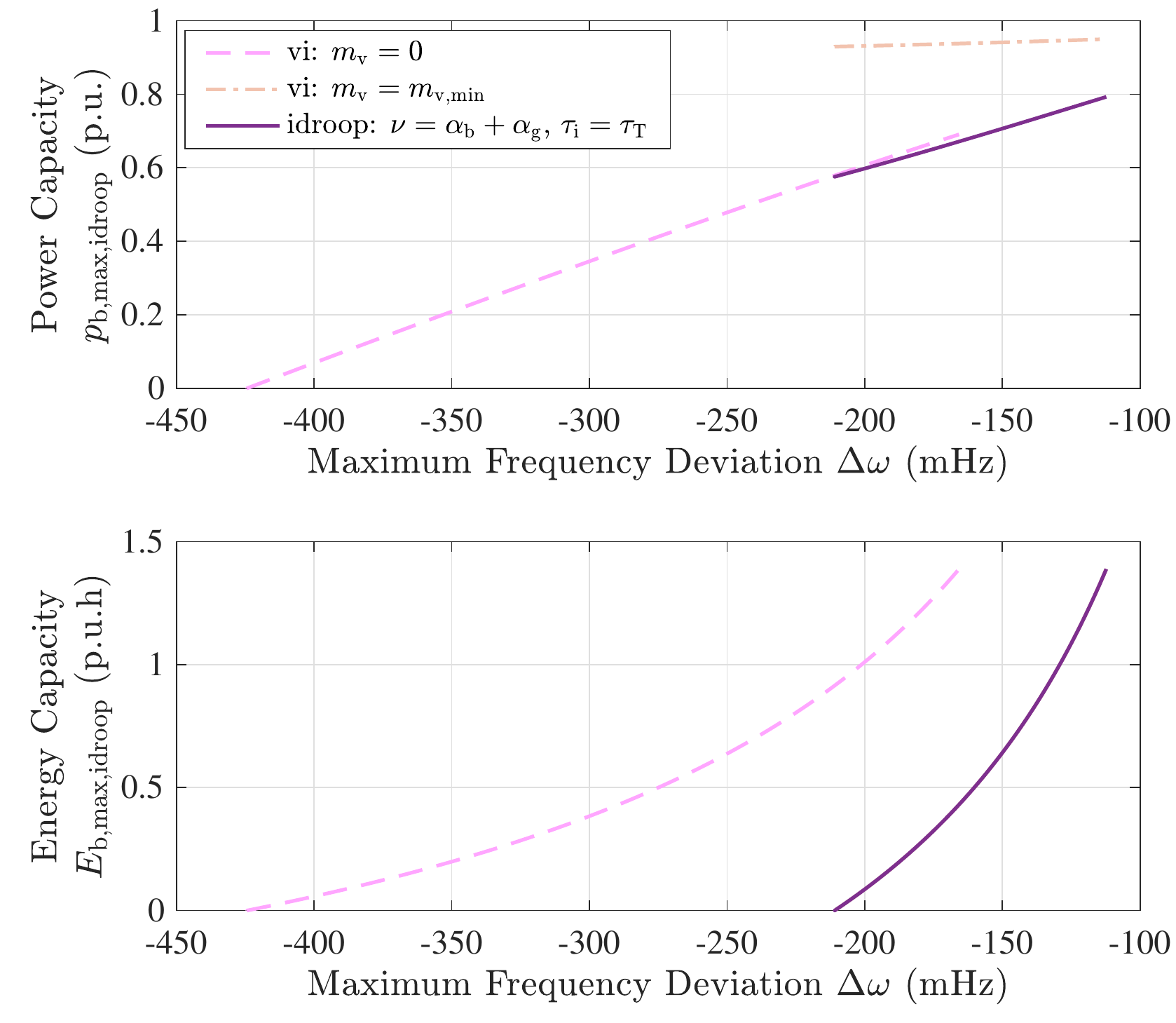}
\caption{Storage capacity requirement under iDroop with $\nu=\alpha_\mathrm{b}+\alpha_\mathrm{g}$ and $\tau_\mathrm{i}=\tau_\mathrm{T}$.}
\label{fig:Pb_Eb-idroop}
\end{figure}

Fig.~\ref{fig:Pb_Eb-idroop} illustrates the storage power and energy requirements for iDroop and virtual inertia control in a slightly different way. This time, instead of setting $\alpha_\mathrm{b}$ to zero, we allow it to vary in order to limit the maximum frequency deviation (in the case of iDroop and virtual inertia with Nadir elimination tuning, it is the same as the steady-state frequency). We then plot the required storage power and energy as a function of maximum frequency deviation $\Delta \omega$ for different control strategies, namely, pure droop control ($m_\mathrm{v}=0$), virtual inertia with $m_\mathrm{v}=m_\mathrm{v,min}$, and iDroop with $\nu=\alpha_\mathrm{b}+\alpha_\mathrm{g}$ and $\qquad\tau_\mathrm{i}=\tau_\mathrm{T}$. The top panel shows that the pure droop control is efficient when the constraints on the maximum frequency deviation are not tight. Both iDroop and virtual inertia require substantial storage power to be executed, with the iDroop being superior. The curves for both iDroop and virtual inertia start from the $\Delta \omega$ values corresponding to $\alpha_\mathrm{b}=0$. Both approaches effectively eliminate the Nadir, and the further decrease in $\Delta \omega$ is achieved by increasing $\alpha_\mathrm{b}$. This leads to a significant growth in the required storage energy capacity, as seen from the bottom panel. In this case, since $\alpha_\mathrm{b} \neq 0$, storage units keep participating in balancing power even when the frequency reaches its steady-state value until the secondary control action returns the frequency back to its nominal value. 
As in the case of virtual inertia, for non-negligible $\alpha_\mathrm{b}$, the required storage capacity by iDroop is $E_{\mathrm{b,max}}=\alpha_{\mathrm{b}}/K_\mathrm{I}$.     
% \enrique{Make sure the commented sentence above is fine.}

\subsection{Robustness to Turbine Time Constant}
We now steer from nominal performance towards robustness performance. Particularly, we are interested in how the frequency deviation changes when the parameter values of iDroop are designed based on a model whose turbine time constant is not exactly known. For this, we tune the iDroop control \eqref{eq:tf-idroop} for a certain value of $\tau_\mathrm{i}$ and then vary the turbine time constant $\tau_\mathrm{T}$.    

\begin{figure}[!t]
\centering
\hspace*{0mm}\includegraphics[width=0.8\columnwidth]{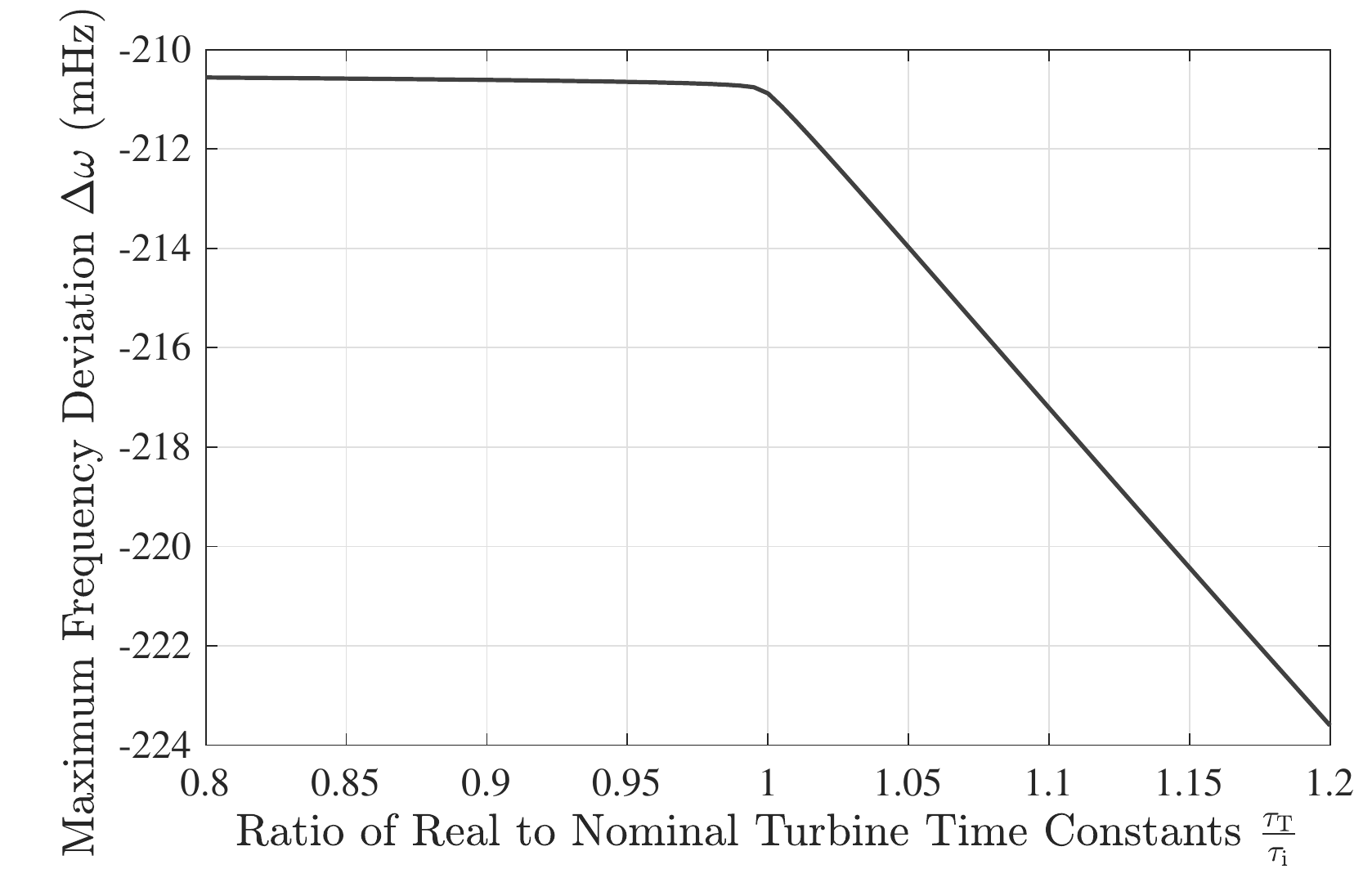}
\caption{Maximum frequency deviation as a function of turbine time constant for iDroop with $\nu=\alpha_\mathrm{b}+\alpha_\mathrm{g}$ and $\tau_\mathrm{i}=\SI{1}{\second}$.}
\label{fig:robust-idroop-tau}
\end{figure}

Fig.~\ref{fig:robust-idroop-tau} shows the sensitivity of the maximum frequency deviation to turbine time constant variation for $\alpha_\mathrm{b}=0$. Notably, if the real turbine time constant is less than the one used by iDroop, i.e., $\tau_\mathrm{T}<\tau_\mathrm{i}$ (the turbine is faster than we expect), then the maximum frequency deviation is the same as predicted. However, if the turbine time constant is greater that the one used by iDroop, i.e., $\tau_\mathrm{T}>\tau_\mathrm{i}$ (the turbine is slower than we expect), then there is a frequency Nadir present and the maximum frequency deviation is greater than we expected with the difference growing proportionally to the increase in the turbine time constant. 

\begin{figure}[!t]
\centering
\hspace*{0mm}\includegraphics[width=0.8\columnwidth]{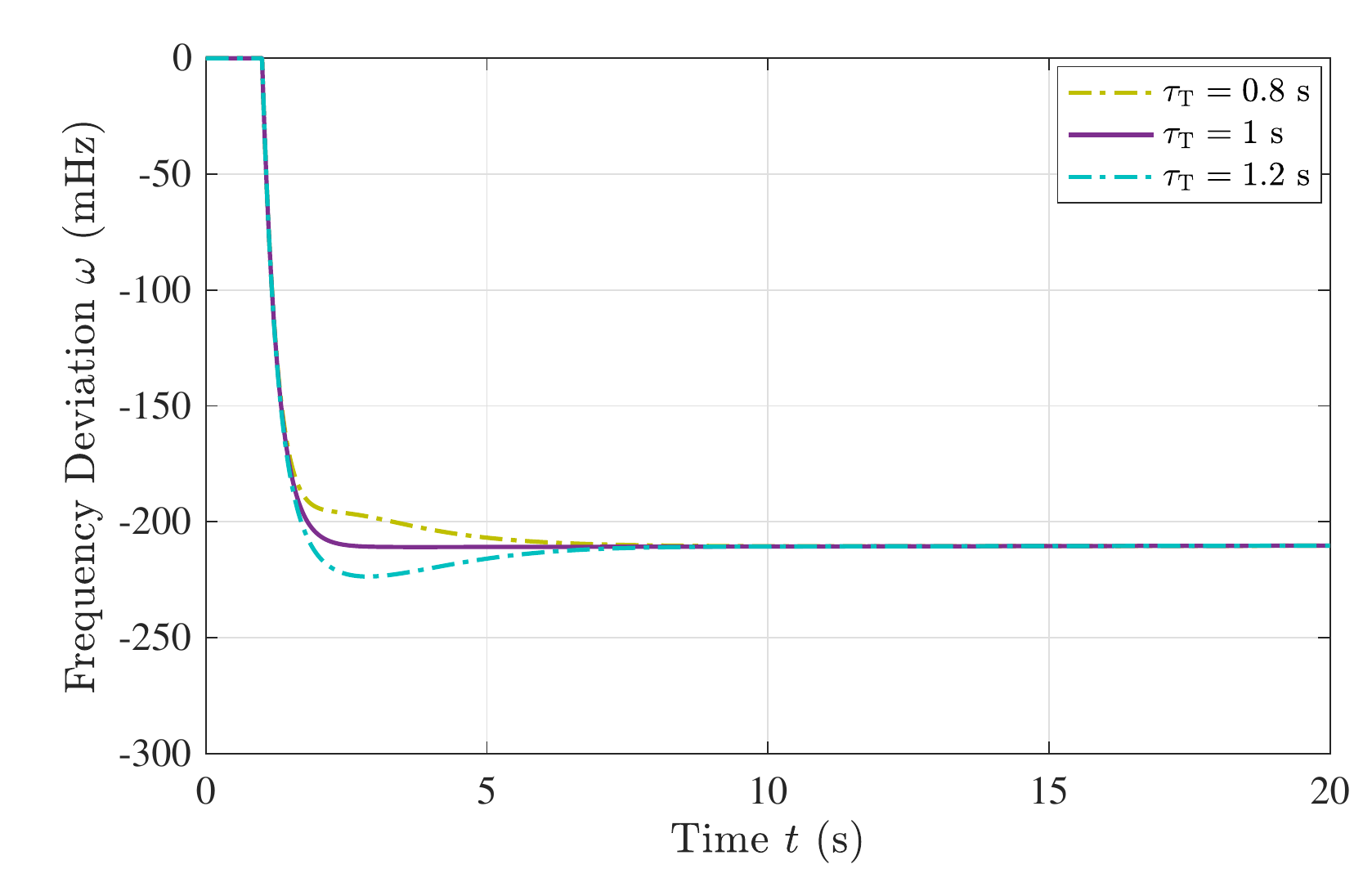}
\caption{Frequency deviation for different values of turbine time constants $\tau_\mathrm{T}$ under iDroop with $\nu=\alpha_\mathrm{b}+\alpha_\mathrm{g}$ and $\tau_\mathrm{i}=\SI{1}{\second}$.}
\label{fig:fre-idroop-tau}
\end{figure}

It is evident from Fig.~\ref{fig:fre-idroop-tau} that iDroop can successfully eliminate Nadir even if $\tau_\mathrm{i} > \tau_\mathrm{T}$, which implies that certain conservative estimates of the turbine time constant can be used. However, this comes at a price of more power and energy capacity requirements from storage, as opposed to the case when the turbine time constant is known accurately.

% \begin{align*}
%     &\hat{\omega} \\=& -\cfrac{\left(\tau_\mathrm{T,real} s + 1\right)\left(s+\delta\right)}{\left[s\left(2H s + \alpha_\mathrm{L}\right)\left(\tau_\mathrm{T,real} s + 1\right)+\left(\alpha_\mathrm{g}s+ K_\mathrm{I}\right)\right]\left(s+\delta\right)+ s\left(\tau_\mathrm{T,real} s + 1\right)\left(\nu s +\delta \alpha_\mathrm{b}\right)}\;.
% \end{align*}

\subsection{Effect of Governor Dead-bands}

We now numerically verify the performance of iDroop in the presence of governor dead-bands, which makes the turbine control non-linear. In the presence of governor dead-bands $\omega_\mathrm{db}$, the turbine only starts reacting when the frequency deviation $\omega$ is outside $\pm \omega_\mathrm{db}$, which is assumed to be \SI{36}{\milli\hertz} (\SI{0.0006}{\pu}). We have: 
$$\varphi_{\omega_\mathrm{db}}(\omega):=
    \begin{cases}
    - \alpha_\mathrm{g}(\omega+\omega_\mathrm{db}) & \omega \leq -\omega_\mathrm{db}\\
    0 & -\omega_\mathrm{db} < \omega < \omega_\mathrm{db}\\
    -\alpha_\mathrm{g}(\omega-\omega_\mathrm{db})  & \omega \geq \omega_\mathrm{db}
    \end{cases}
$$
in \eqref{eq:tur-dyan} instead of $- \alpha_\mathrm{g}\omega$.

\begin{figure}[t!]
\centering
\includegraphics[width=0.8\columnwidth]{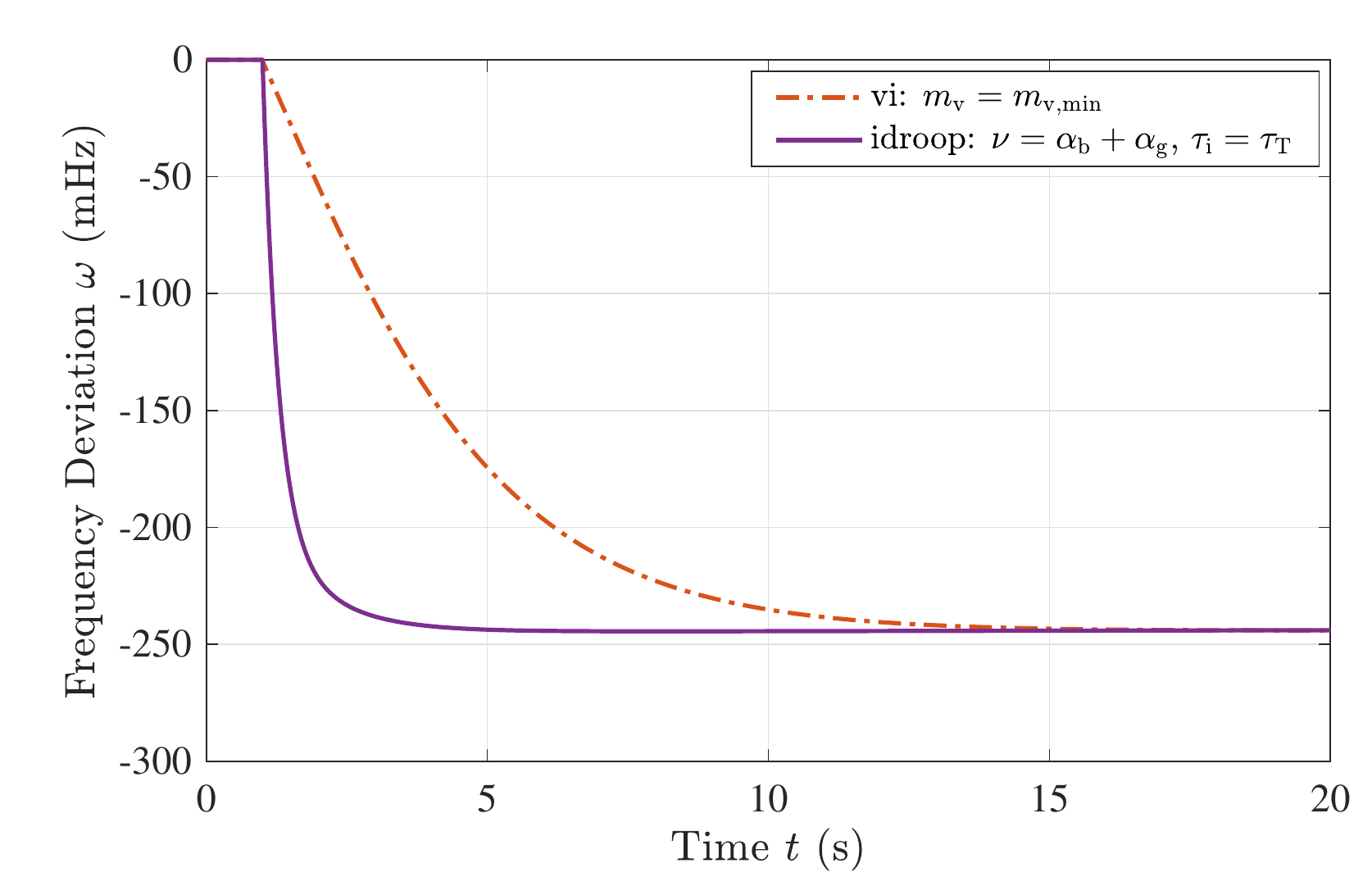}
% \subfigure[Power output from storage]
% {\includegraphics[width=0.95\columnwidth]{Figures/dyn_Pb_idroop_ab_0_F}}\label{fig:dyn-Eb_Pb-idroop-ab0}
% \subfigure[Power output from turbine]
% {\includegraphics[width=0.95\columnwidth]{Figures/dyn_Pm_idroop_ab_0_F}}\label{fig:dyn-Eb_Pb-idroop-ab0}
% \subfigure[Energy supply from storage]
% {\includegraphics[width=0.95\columnwidth]{Figures/dyn_Eb_idroop_ab_0_F}}\label{fig:dyn-Eb_Pb-idroop-ab0}
\caption{Frequency deviations under virtual inertia and iDroop control with Nadir elimination tuning for a step
power imbalance with \SI{\pm 36}{\milli\hertz} governor dead-bands considered.}
\label{fig:dyn-db-ab0}
\end{figure}

Fig.~\ref{fig:dyn-db-ab0} compares how the Nadir elimination tuning of virtual inertia and iDroop performs when the system, with \SI{\pm 36}{\milli\hertz} governor dead-bands, suffers from a step power imbalance for $\alpha_\mathrm{b} = 0$. Although the presence of dead-bands tends to lower the steady-state frequency to some degree, it has a minimal (if any) effect on the Nadir elimination performance of both virtual inertia and iDroop.

%----------------------------------------------------%
%               S E C T I O N  V
%----------------------------------------------------%  

\section{Conclusions and future work}\label{Sec:conc}

We have presented an efficient strategy for frequency control with energy storage that is capable of eliminating the frequency Nadir following power disturbances. Such a control can be especially effective for providing frequency security to modest-sized power grids or microgrids, which are known to suffer from excessively large frequency Nadirs. Our approach is based on iDroop control, where the storage response to frequency is tuned to make the system dynamics become effectively first-order. We have demonstrated that the iDroop approach is much more efficient than both regular droop control and virtual inertia from storage. We have tested our method on a rather simple, yet representative model of a single-area power grid with first-order turbine dynamics of a combined generator. The future research will need to concentrate on such important extensions, including multiple-area systems, higher order turbine and generator models, system robustness in response to frequency measurement noise and delays. Finally, explicit consideration of the storage inverter dynamics, especially the action of PLL (phase-locked loop) for frequency measurements will be done to provide the final verification of the practical performance of the developed method.

%\todoinr{for larger $\alpha_b$, it takes a extremely long time for power and energy plots to reach steady-state such that the figures are not that nice, that's why I didn't include them here}

% \begin{figure}[!t]
% \centering
% \hspace*{0mm}\includegraphics[width=0.95\columnwidth]{Figures/dyn_fre_idroop_ab_0}
% \caption{Comparison of transient responses under virtual inertia and iDroop to a \SI{1}{\pu} step
% load perturbation of the one-area model with \SI{\pm 36}{\milli\hertz} governor dead-bands considered for $\alpha_\mathrm{b} = \SI{0}{\pu}$.}
% \label{fig:dyn-fre-idroop-ab0}
% \end{figure}

% \begin{figure}[!t]
% \centering
% \hspace*{0mm}\includegraphics[width=0.95\columnwidth]{Figures/dyn_Pb_Eb_idroop_ab_0}
% \caption{Comparison of transient responses under virtual inertia and iDroop to a \SI{1}{\pu} step
% load perturbation of the one-area model with \SI{\pm 36}{\milli\hertz} governor dead-bands considered for $\alpha_\mathrm{b} = \SI{0}{\pu}$.}
% \label{fig:dyn_fre-idroop-ab0}
% \end{figure}

\bibliographystyle{IEEEtran}
\bibliography{PSCC2020}

\end{document}